\documentclass[11pt]{article}
\usepackage{putex}
\usepackage{graphicx}
\usepackage{latexsym,amsmath,amsfonts,amssymb}
\usepackage{bbm}
\usepackage{cite}
\usepackage{mathtools}
\renewenvironment{thebibliography}[1]{%
\begin{oldthebibliography}{#1}%
\setlength\itemsep{-2mm}%
}%
{%
\end{oldthebibliography}%
}
\newcommand{\dvol}{d\mathrm{vol}}

\newcommand{\ie}{\textit{i.e.}}

\numberwithin{equation}{section}

\newcommand{\be}{\begin{equation}} \newcommand{\ee}{\end{equation}}
\newcommand{\bea}{\begin{equation} \begin{aligned}} \newcommand{\eea}{\end{aligned} \end{equation}}

\newcommand{\fg}{\mathfrak{g}}

\newcommand{\fz}{\mathfrak{z}}
\newcommand{\unit}{\mathbbm{1}}

\DeclareMathOperator{\Tr}{Tr}

\DeclareMathOperator{\rank}{rank}
\DeclareMathOperator*{\Res}{Res}

\begin{document}

%%%%%%% title page %%%%%%%%%

\title{Higgs branch localization of $\mathcal{N}=1$ theories on $S^3\times S^1$}

\authors{Wolfger Peelaers$^\spadesuit$}

\institution{YITP}{${}^\spadesuit$
C.N. Yang Institute for Theoretical Physics, Stony Brook University,  \cr
$\;\;\,$ Stony Brook, NY 11794, USA}

\abstract{We apply the method of Higgs branch localization to the $\mathcal{N}=1$ supersymmetric partition function on $S^3\times S^1.$ As a result, we show that it can be written as the product of an elliptic vortex and anti-vortex partition function summed over a finite number of Higgs vacua. }

\preprint{YITP-SB-14-06}

\maketitle

%%%%%% end of title page %%%%%%

%%%%%% table of contents %%%%%%
{%\small
\setcounter{tocdepth}{2}
\setlength\parskip{-0.7mm}
\tableofcontents
}
%%%%%%%%%%%%%%%%%%%%%

\section{Introduction and outlook}\label{sec: intro}
The application of localization techniques \cite{Witten:1988ze,Witten:1991zz} to the computation of the partition function of supersymmetric gauge theories on compact Euclidean manifolds and to the evaluation of expectation values of supersymmetric observables in the same class of theories, pioneered in \cite{Pestun:2007rz}, has led to an impressive number of exact results in various dimensions, see for example \cite{Benini:2012ui,Doroud:2012xw,Gadde:2013dda,Benini:2013nda,Benini:2013xpa,Kim:2009wb,Imamura:2011su,Kapustin:2009kz,Jafferis:2010un,Hama:2010av,Hama:2011ea,Imamura:2011wg,Alday:2012au,Alday:2013lba,Nawata:2011un,Gomis:2011pf,Hama:2012bg,Closset:2013sxa,Kallen:2012cs,Hosomichi:2012ek,Kim:2012ava,Imamura:2012xg} for a non-exhaustive list in dimensions two to five.

Most of these exact results are obtained using the so-called Coulomb branch localization, and the result then takes the form of a matrix integral over the classical Coulomb branch, \ie\  the localization locus consists of constant vector multiplet scalars or holonomies around circles. The integrand is given by the product of a classical action, a one-loop piece describing the quadratic fluctuations around the localization locus and possibly some non-perturbative contributions. Recently, an alternative localization method has been developed in two dimensions \cite{Benini:2012ui,Doroud:2012xw}, called Higgs branch localization, where the localization locus is a finite number of Higgs vacua each supporting an infinite tower of vortex and anti-vortex configurations. Technically, it can be achieved by considering an alternative deformation term or equivalently by modifying the integration contour in complexified field space in the path integral. The result now takes the schematic form of a sum over the aforementioned Higgs vacua, with as its summand the product of a classical action, a one-loop piece evaluated on the Higgs vacua and the product of a vortex and an anti-vortex partition function. Such a factorized expression was actually first observed in three dimensions in \cite{Pasquetti:2011fj} by manipulating the matrix integral computing the partition function on the squashed three-sphere, and subsequently understood in terms of holomorphic blocks in \cite{Beem:2012mb} and by deforming the geometry in \cite{Alday:2013lba}, and generalized to larger gauge groups and $S^2\times S^1$ in \cite{Taki:2013opa,Krattenthaler:2011da,Hwang:2012jh}. These three-dimensional factorization results were explained from the point of view of Higgs branch localization in \cite{Fujitsuka:2013fga,Benini:2013yva}. 

It is now a natural question to ask if Higgs branch localization can be applied to four-dimensional theories as well. This would imply that the partition function can be factorized. In this note, we address this question -- and answer it positively -- for $\mathcal{N}=1$ supersymmetric gauge theories on $S^3\times S^1.$\footnote{Recently the gravity dual of these theories has been studied in \cite{Cassani:2014zwa}.} At an RG fixed point, the partition function on this geometry is known to describe the superconformal index \cite{Kinney:2005ej}, which encodes information about the protected spectrum of the corresponding flat space theory. A prescription to write down the Coulomb branch expression computing the index of the IR fixed point to which a given Lagrangian UV theory flows, was first given by R\"omelsberger in \cite{Romelsberger:2005eg,Romelsberger:2007ec} and it takes the form of a matrix integral over the holonomy along the temporal circle $S^1$ of the one-loop determinants, which are typically expressed in terms of the plethystic exponential of the single letter partition functions of the fields in the UV theory, but can also be written in terms of elliptic hypergeometric functions \cite{Dolan:2008qi}. Our main result shows that the index can alternatively be written as
\begin{equation}
I = \sum_{\text{Higgs vacua}} Z_{\text{cl}}\  Z_{\text{1-loop}}^\prime\  Z_{\text{v}}\  Z_{\text{av}}\;,
\end{equation}
which has the typical form of a Higgs branch localized result. Here $Z_{\text{v}}$ and $Z_{\text{av}}$ are the contributions from vortex-membranes wrapping a torus at two distinct points in the geometry. As such they are given  in terms of the elliptic uplift of the usual vortex partition function in the $\Omega$ background \cite{Shadchin:2006yz}.

The superconformal index is a powerful tool in checking various dualities, see for example \cite{Dolan:2008qi,Spiridonov:2009za,Spiridonov:2011hf}. It would be very interesting to study the effects of such dualities on the vortex-membrane partition function. Moreover, the elliptic vortex partition functions we encounter are naturally expected to have nice modular properties. It would be interesting to study these alongside the modular properties of the index itself \cite{Razamat:2012uv}. The factorization results obtained in this paper are expected to be just one instance of a rich structure involving the four-dimensional uplift of the holomorphic blocks of \cite{Beem:2012mb}. Unraveling this structure is an outstanding problem. Finally, the $\mathcal{N}=1$ index can be further decorated with surface operators. Three possible approaches can be used to introduce them, namely to construct them as the IR limit of vortex configurations as in \cite{Gaiotto:2012xa}, to perform a localization computation as in \cite{Drukker:2012sr} for vortex-loops, or to consider a coupled 2d-4d system as in \cite{Gadde:2013dda}. Their connections among each other and with the vortex factorization should be study-worthy.

On the other hand, we expect the techniques used in this paper to be applicable to $\mathcal{N}=1$ supersymmetric theories on different geometries as well, most obviously $L(r,1)\times S^1$\cite{Benini:2011nc,Razamat:2013opa}, but also more generally in theories with more supersymmetry. For example, the $\mathcal{N}=2$ superconformal index for theories of class $\mathcal{S}$ is computed by a TQFT correlator \cite{Gadde:2009kb,Gadde:2011uv} and it would be very interesting to study its interplay with a possible vortex anti-vortex factorization. We leave this for future work.

The outline of this note is as follows. In section \ref{section_KS} we introduce the index we want to compute and construct the deformed background on which the computation of the partition function achieves that goal. Next, we derive the BPS equations in section \ref{section_BPSequations} and find various classes of solutions to them in section \ref{section_BPSsolutions}. We compute the index on the various solutions in section \ref{section_computationindex}. In section \ref{section_mathcingCB} we match our Higgs branch expression with the Coulomb branch expression in some examples by manipulating the matrix integral. Here we also explain how our results apply in the absence of an abelian factor in the gauge group. Finally, the appendices contain the spinor conventions we use, the $\mathcal{N}=1$ algebra and some useful identities satisfied by the elliptic gamma function.

\

\emph{Note added.} When this work was under completion, we became aware of \cite{Yoshida:2014qwa} which has some overlap with our paper.

\section{Killing spinors on $S^3\times S^1$, supersymmetric index and deformed background}\label{section_KS}
$\mathcal{N}=1 $ supersymmetric theories on $S^3\times \mathbb{R}$ were explicitly constructed in \cite{Sen:1985ph} and later also in \cite{Romelsberger:2005eg}. A systematic study of supersymmetric theories on Euclidean four-manifolds, among which $S^3\times S^1$,  with four or less supercharges was performed in \cite{Festuccia:2011ws,Dumitrescu:2012ha,Dumitrescu:2012at} by considering the rigid limit of supergravity. Their method constructs supersymmetric backgrounds as solutions to the Killing spinor equation, which in turn is obtained by setting to zero the gravitino variations, as well as -- in the presence of flavor symmetries -- to the equations that set to zero the gaugino variations, while treating the bosonic auxiliary fields as arbitrary background fields. \footnote{The parameter dependence of partition functions on these four-dimensional backgrounds was studied in \cite{Closset:2013vra}.}

For the particular case of the index, it is illustrative to construct the sought-after supersymmetric background differently, namely by turning on background gauge fields associated to the charges appearing in the supersymmetric index such that in a path integral formulation they have the effect of precisely implementing the twisted boundary conditions along the temporal circle imposed by the index. As a preliminary step, we first construct the solutions to the conformal Killing spinor equations on $S^3\times \mathbb{R}$ and select the Killing spinor associated to the supercharge with respect to which we will compute the index. Its lack of periodicity along $\mathbb{R}$ is remedied by the twisted boundary conditions imposed by the associated index.

\paragraph{Killing spinors on $S^3\times \mathbb{R}$} We would like to solve the Killing spinor equation\footnote{Our spinor conventions are summarized in appendix \ref{section:appendix_spinorconventions}.}
\begin{equation}\label{KSequation}
D_\mu \varepsilon = \left(\partial_\mu +\frac{1}{4}\omega_\mu^{\ mn}\gamma_{mn} -i V_\mu\gamma_5\right)\varepsilon = \gamma_\mu \tilde\varepsilon
\end{equation}
on $S^3\times \mathbb{R}$ with metric
\begin{equation}
ds^2 = ds^2_{S^3} + d\tau^2 = \ell^2\left( d\theta^2 + \cos^2\theta\  d\varphi^2 + \sin^2\theta\  d\chi^2 \right) + d\tau^2\;.
\end{equation}
Upon choosing the vielbeins $e^1 = \ell \cos\theta\ d\varphi,$ $e^2 = \ell \sin\theta\ d\chi,$ $e^3 = \ell d\theta,$ $e^4 = d\tau,$ one finds the non-zero components of the spin connection to be $\omega^{13} = -\sin\theta\ d\varphi$ and $\omega^{23}=\cos\theta d\chi.$ At this point we also set the $U(1)_R$ background field $V_\mu$ to zero.

Our first step in solving \eqref{KSequation} is to write $\tilde \varepsilon = \gamma^4 \hat\varepsilon$ and decompose $\varepsilon$ and $\hat\varepsilon$ into their right and left-handed components, which we denote as $\varepsilon = \left( \begin{smallmatrix} \eta \\ \zeta \end{smallmatrix}\right)$ and similarly $\hat\varepsilon = \left( \begin{smallmatrix} \hat \eta \\ \hat \zeta \end{smallmatrix}\right).$ The equation \eqref{KSequation} then splits as
\begin{eqnarray}
\left(\partial_\mu +\frac{1}{4}\omega_\mu^{\ mn}\sigma_{mn} \right)\eta &=& -i\sigma_\mu \hat\eta \\
\left(\partial_\mu +\frac{1}{4}\omega_\mu^{\ mn}\bar\sigma_{mn} \right)\zeta &=& i\bar\sigma_\mu \hat\zeta\;.
\end{eqnarray}
Next, making a factorized Ansatz $\eta = f(\tau)\ \eta_{S^3}$ and $\hat\eta = f(\tau)\ \hat\eta_{S^3},$ where $\eta_{S^3}$ and $\hat \eta_{S^3}$ only depend on the coordinates on the three-sphere, one immediately recognizes that the spatial part of the Killing spinor equation simplifies to the Killing spinor equation on $S^3$ 
\begin{equation}
\left(\partial_{\hat{\mu}} +\frac{1}{4}\omega_{\hat{\mu}}^{\ \hat{m}\hat{n}}\sigma_{\hat{m}\hat{n}} \right)\eta_{S^3}^{(s_1,t_1)} = -i\sigma_{\hat\mu} \hat\eta_{S^3}^{(s_1,t_1)} \;,
\end{equation}
where $\hat\mu = \varphi, \chi, \theta.$ Its solutions are given by \cite{Hama:2011ea}
\begin{equation}
\eta_{S^3}^{(s_1,t_1)} = \begin{pmatrix}
e^{\frac{i}{2}(s_1\chi+t_1\varphi-s_1t_1\theta)}\\
-s_1\ e^{\frac{i}{2}(s_1\chi+t_1\varphi+s_1t_1\theta)}
\end{pmatrix}\;,\qquad (s_1,t_1 = \pm)
\end{equation}
if $\hat\eta_{S^3}^{(s_1,t_1)} = \frac{s_1t_1}{2\ell}\eta_{S^3}^{(s_1,t_1)}.$ The time dependence is then determined by $\partial_\tau f(\tau) = \frac{s_1t_1}{2\ell}f(\tau)$ which implies that $f(\tau) = e^{\frac{s_1t_1}{2\ell}\tau}.$ In total one thus finds that
\begin{equation}
\eta^{(s_1,t_1)} = e^{\frac{s_1t_1}{2\ell}\tau} \eta_{S^3}^{(s_1,t_1)}\;, \qquad \hat\eta^{(s_1,t_1)} = \frac{s_1t_1}{2\ell}\eta^{(s_1,t_1)}\;.
\end{equation}
Similarly, one finds that
\begin{equation}
\zeta^{(s_2,t_2)} = e^{-\frac{s_2t_2}{2\ell}\tau} \zeta_{S^3}^{(s_2,t_2)}\;, \qquad \hat\zeta^{(s_2,t_2)} = -\frac{s_2t_2}{2\ell}\zeta^{(s_2,t_2)}\;,
\end{equation}
and the most general four-component solution is thus
\begin{equation}\label{KSsolution}
\varepsilon = \sum_{s_1,t_1=\pm} a_{s_1,t_1}\begin{pmatrix}
\eta^{(s_1,t_1)} \\ 0
\end{pmatrix} + \sum_{s_2,t_2=\pm} b_{s_2,t_2}\begin{pmatrix}
0 \\ \zeta^{(s_2,t_2)}
\end{pmatrix}
\end{equation}
We found eight independent supercharges as expected for a superconformal $\mathcal{N}=1$ theory. Note that the Killing spinors are not periodic along the time circle which signals the need of twisted boundary conditions.

\paragraph{Killing spinors for supersymmetric index}
We choose the combination of supercharges described by the four-component spinor $\varepsilon$ in \eqref{KSsolution} with as only non-zero coefficients $a_{++}=1$ and $b_{--}=1,$ 
\begin{equation}\label{KSsol1}
\varepsilon_1 = \begin{pmatrix}
\eta^{(+,+)} \\ \zeta^{(-,-)}
\end{pmatrix}\;.
\end{equation}
It satisfies $D_\mu \varepsilon_1 = \frac{1}{2\ell}\gamma_\mu \gamma_4 \gamma_5 \varepsilon_1\;.$ The bilinears appearing in the algebra (see formula \eqref{algebrabilinears} in appendix \ref{section_appendixalgebra}) are then given by
\begin{equation}
v_1^\mu \partial_\mu = \frac{2}{\ell}\left(-i(\partial_\varphi + \partial_\chi) + \ell \partial_\tau \right)\;, \qquad \rho_1 = 0\;, \qquad \alpha_1 = \frac{3i}{\ell}\;,
\end{equation}
which upon plugging in \eqref{algebradeltasq} result in
\begin{align}
\delta^2_{\varepsilon_1} = -\frac{2}{\ell}\left(-\ell \mathcal{L}^A_{\partial_\tau} + i \mathcal{L}^A_{\partial_\varphi + \partial_\chi} +\frac{3}{2} R \right)\;.
\end{align}
Introducing the operators 
\begin{equation}
\Delta = -\ell \mathcal{L}^A_{\partial_\tau}\;, \qquad j_1 = -\frac{i}{2} \mathcal{L}^A_{\partial_\chi + \partial_\varphi}\;, \qquad j_2 = -\frac{i}{2} \mathcal{L}^A_{\partial_\chi - \partial_\varphi}\;,
\end{equation}
one can also write
\begin{align}
\delta^2_{\varepsilon_1} = -\frac{2}{\ell}\left(\Delta - 2 j_1 + \frac{3}{2}R \right)\;.
\end{align}

The action of the operators $\Delta, j_1, j_2,$ and $R$ on $\varepsilon_1$ is given by
\begin{equation}
\Delta \epsilon_1 = -\frac{1}{2}\gamma_5\varepsilon_1\;, \qquad j_1 \epsilon_1 = \frac{1}{2}\gamma_5\varepsilon_1\;, \qquad j_2 \epsilon_1 = 0\;, \qquad R \epsilon_1 = \gamma_5\varepsilon_1\;.
\end{equation}
Note that as expected $\left(\Delta - 2 j_1 + \frac{3}{2}R\right)\varepsilon_1 = 0.$ We can find two more linearly independent charges that vanish on the Killing spinor, namely $2j_1-R$ and $j_2.$

Alternatively, we could choose the combination of supercharges described by the four-component spinor with as only non-zero coefficients $a_{-+}=1$ and $b_{+-}=1$ 
\begin{equation}
\varepsilon_2 = \begin{pmatrix}
\eta^{(-,+)} \\ \zeta^{(+,-)}
\end{pmatrix}\;,
\end{equation}
which satisfies $D_\mu \varepsilon_2 = -\frac{1}{2\ell}\gamma_\mu \gamma_4 \gamma_5 \varepsilon_2\;.$ Then we find
\begin{align}
\delta^2_{\varepsilon_2} = \frac{2}{\ell}\left(-\ell \mathcal{L}^A_{\partial_\tau} + i \mathcal{L}^A_{\partial_\chi - \partial_\varphi} -\frac{3}{2} R \right) = \frac{2}{\ell}\left(\Delta -2j_2 -\frac{3}{2} R \right)\;.
\end{align}
Now one has $\Delta \epsilon_2 = \frac{1}{2}\gamma_5\varepsilon_2,$ $j_1 \epsilon_2 = 0,$ $j_2 \epsilon_2 = -\frac{1}{2}\gamma_5\varepsilon_2,$ and $R \epsilon_2 = \gamma_5\varepsilon_2.$
We find three linearly independent charges that vanish on the Killing spinor, $\Delta - 2 j_2 - \frac{3}{2}R,$ $2j_2+R$ and $j_1.$

\paragraph{Supersymmetric index and deformed background}
One can introduce two inequivalent superconformal indices in $\mathcal{N}=1$ theories, a left-handed one and a right-handed one, namely
\begin{equation}
\begin{array}{c}
I_1(t,y,\zeta_j) = \Tr (-)^F\  e^{-\beta \left(\Delta - 2j_1 + \frac{3}{2}R\right)} \  t^{3(2j_1 - R)}\  y^{2j_2}\ \prod_j\zeta_j^{F_j}\\
I_2(t,y,\zeta_j) = \Tr (-)^F\  e^{-\beta \left(\Delta - 2j_2 - \frac{3}{2}R\right)}\  t^{3(2j_2 + R)}\  y^{2j_1}\ \prod_j\zeta_j^{F_j}
\end{array}
\quad \text{where}\enskip t=e^{-\xi}, y=e^{i\eta} \text{ and } \zeta_j=e^{i\fz_j},
\end{equation}
where $\beta \ell$ is the circumference of the temporal circle and $F_j$ are the Cartan generators of the flavor symmetry group. Convergence requires that $|t|<1.$ These indices are precisely computed with respect to the charges described by the Killing spinors $\varepsilon_1$ and $\varepsilon_2$ respectively. It is very important to remark that all charges appearing in the index need to be non-anomalous -- we will always assume this to be the case. From here onward, we will focus on the index $I_1,$ knowing that $I_2$ can be dealt with completely similarly. 

In the path integral formulation, the insertion of the chemical potentials in the trace leads to twisted boundary conditions on all fields
\begin{equation}
\Phi(\tau + \beta \ell) = e^{\beta \left(- 2j_1 + \frac{3}{2}R\right)}\ t^{-3(2j_1 - R)}\  y^{-2j_2}\  \prod_j\zeta_j^{-F_j}\  \Phi(\tau)\;,
\end{equation}
which are indeed also the boundary conditions satisfied by the Killing spinor $\varepsilon_1.$ Alternatively, one can turn on flat background gauge connections along the temporal circle
\begin{equation}\label{bgfields}
V_\mu = \left(0,0,0,i\left( \frac{3\beta - 6\xi}{2\beta\ell}\right)\right)\;, \qquad \tilde V^{(j)}_\mu = \left(0,0,0,\frac{\fz_j}{\beta \ell} \right)\;,
\end{equation}
for the R-symmetry and the flavor symmetry respectively. The twists by the rotational charges $j_1$ and $j_2$ furthermore impose the identification
\begin{equation}
(\varphi, \chi, \theta, \tau ) \sim \left(\varphi +\frac{i}{2}(-2\beta + 6\xi + 2i\eta) , \chi +\frac{i}{2}(-2\beta + 6\xi - 2i\eta), \theta, \tau+ \beta \ell\right)\;.
\end{equation}
Introducing the coordinates 
\begin{equation}
\hat\varphi = \varphi -\frac{i}{2}(-2\beta + 6\xi + 2i\eta)\frac{\tau}{\beta\ell} \;,\quad \hat\chi = \chi -\frac{i}{2}(-2\beta + 6\xi - 2i\eta)\frac{\tau}{\beta\ell}\;,\quad\hat\theta = \theta\;,\quad \hat\tau = \tau\;,
\end{equation}
the identification simplifies to $(\hat\varphi, \hat\chi, \hat\theta, \hat\tau ) \sim (\hat\varphi , \hat\chi , \hat\theta, \hat\tau+ \beta \ell)\;.$ The metric in these hatted coordinates reads
\begin{eqnarray}\label{deformedmetric}
ds^2 &=&   \ell^2\cos^2\hat\theta\  \left(d\hat\varphi+\frac{i}{2\beta\ell}(-2\beta + 6\xi + 2i\eta)\ d\hat\tau \right)^2 + \notag \\
& +& \ell^2\sin^2\hat\theta\  \left(d\hat\chi+\frac{i}{2\beta\ell}(-2\beta + 6\xi - 2i\eta)\ d\hat\tau\right)^2 + \ell^2 d\hat\theta^2 +d\hat\tau^2 \;,
\end{eqnarray}
and is complexified. Its vielbeins are
\begin{eqnarray}
e^1 &=  \ell \cos\hat\theta\  \left(d\hat\varphi+\frac{i}{2\beta\ell}(-2\beta + 6\xi + 2i\eta)\ d\hat\tau \right)\;,\qquad  & e^3 = \ell d\hat\theta\;,\\
e^2 &= \ell \sin\hat\theta\  \left(d\hat\chi+\frac{i}{2\beta\ell}(-2\beta + 6\xi - 2i\eta)\ d\hat\tau\right)\;, \qquad & e^4 = d\hat\tau\;,
\end{eqnarray}
while the dual frame vectors are given by
\begin{eqnarray}
e_1 &=&  \left(\ell \cos\hat\theta\right)^{-1}\  \partial_{\hat\varphi}\;, \qquad e_2 = \left(\ell \sin\hat\theta\right)^{-1}\ \partial_{\hat\chi}\;,\qquad e_3 = \ell^{-1} \partial_{\hat\theta}\;, \\
e_4 &=& \partial_{\hat{\tau}} - \frac{i}{2\beta\ell}(-2\beta + 6\xi + 2i\eta)\  \partial_{\hat\varphi}- \frac{i}{2\beta\ell}(-2\beta + 6\xi - 2i\eta)\  \partial_{\hat\chi}\;,
\end{eqnarray}
and the non-zero components of the spin connection read
\begin{eqnarray}
\omega^{13} &=& -\sin\hat\theta\ \left(d\hat\varphi+\frac{i}{2\beta\ell}(-2\beta + 6\xi + 2i\eta)\ d\hat\tau \right)\\\
\omega^{23}&=&\cos\hat\theta\  \left(d\hat\chi+\frac{i}{2\beta\ell}(-2\beta + 6\xi - 2i\eta)\ d\hat\tau\right)\;.
\end{eqnarray} 
The solution to the Killing spinor equation $D_\mu \varepsilon = \gamma_\mu \tilde\varepsilon$ on the deformed background, corresponding to $\varepsilon_1$ in \eqref{KSsol1}, is given by
\begin{equation}
\varepsilon_1 = \begin{pmatrix}
\eta^{(+,+)}_{S^3} \\ \zeta^{(-,-)}_{S^3}
\end{pmatrix}\;,
\end{equation}
and satisfies $D_\mu \varepsilon_1 = \frac{1}{2\ell}\gamma_\mu \gamma_4 \gamma_5 \varepsilon_1\;.$ The square of the supersymmetry variation equals
\begin{equation}\label{Qsquared}
\delta^2_{\varepsilon_1}= -\frac{2}{\ell}\left[ -\ell\ \mathcal{L}^A_{\partial_{\hat\tau}} + \frac{6i\xi}{2\beta}\ \mathcal{L}^A_{\partial_{\hat\varphi} + \partial_{\hat\chi}} + \frac{\eta}{\beta}\ \mathcal{L}^A_{\partial_{\hat\chi} - \partial_{\hat\varphi}}  + \frac{3\xi}{\beta} R + \frac{i}{\beta}\sum_j \fz_j F_j \right]\;.
\end{equation}

Thanks to pairwise cancellation, the index only receives contributions from states satisfying $\delta_{\varepsilon_1}^2=0.$ It is thus independent of the parameter $\beta,$ and it will be convenient to choose it such that the metric \eqref{deformedmetric} is real, namely $\beta = 3\xi.$ From now on, we make this choice for $\beta$ and further omit the hats.

\paragraph{Fayet-Iliopoulos term}
It is well known that both the gauge and the matter Lagrangian are $\mathcal{Q}$-exact.\footnote{We use $\delta_{\epsilon_1}$ and $\mathcal{Q}$ interchangeably.} However, if the gauge group contains an abelian factor\footnote{In the presence of an abelian factor, the theory develops a Landau pole. However, as was also argued in \cite{Gaiotto:2012xa}, one can exploit the independence of the index on the gauge coupling to suppress the Landau pole arbitrarily by making the gauge coupling smaller and smaller. }, we can write down a Fayet-Iliopoulos term \cite{Romelsberger:2005eg}. Indeed, if the Killing spinor satisfies $D_\mu \varepsilon_1 = \frac{1}{2\ell}\gamma_\mu\gamma_4\gamma_5\varepsilon_1,$ then it is easy to convince oneself that $\delta_{\varepsilon_1}(D+\frac{2}{\ell} A_4) = D_\mu(\bar\varepsilon_1 \gamma_5\gamma^\mu \lambda)\;.$ When integrated over the compact space, the variation of $D+\frac{2}{\ell} A_4$ vanishes and thus results in an invariant action. Note however, that in order for the action to be invariant under large gauge transformations along the 4-direction the properly normalized FI parameter needs to be an integer. Due to its discrete nature it avoids the common lore that the index does not depend on continuous parameters.

\section{The BPS equations}\label{section_BPSequations}
The BPS equations for the vectormultiplet of gauge group $G$ are obtained by setting to zero the gaugino variation
\begin{equation}
0=\delta_{\varepsilon_1} \lambda = -\frac{1}{2}\gamma^{\mu\nu} F_{\mu\nu}\ \varepsilon_1 -  \gamma_5 \ D\  \varepsilon_1\;.
\end{equation}
Upon solving the resulting four equations for $F_{14},F_{24},F_{34},D$ one obtains
\begin{eqnarray}\label{generalBPSeqnYM1}
F_{14} &= i \sin\theta\  F_{12}\;,\qquad\qquad F_{34} &= -i \left(\cos\theta\  F_{13} + \sin\theta\  F_{23}\right)\;, \\  \label{generalBPSeqnYM2}
F_{24} &= -i \cos\theta\  F_{12}\;,\qquad \qquad D &= i\cos\theta\  F_{23} - i \sin\theta\ F_{13} \;.
\end{eqnarray}
Declaring that all fields are real, immediately leads to the localization locus $F_{\mu\nu}=D=0.$ Flat connections on $S^3\times S^1$ are given by $A = \frac{a}{3\xi \ell} d\tau\;,$ for arbitrary holonomy $a.$ Alternatively, we can obtain the localization equations as the zero-locus of the bosonic part of the deformation action 
\begin{equation}
\mathcal{L}_{\text{YM}}^{\text{def}} = \frac{1}{4} \mathcal Q\ \Tr\  (\mathcal Q \lambda)^\ddagger \lambda,
\end{equation}
where the action of the formal hermitian conjugate $\ddagger$ operator on $\mathcal Q \lambda$ is
\begin{equation}
(\mathcal Q \lambda)^\ddagger =  \frac{1}{2} \varepsilon_1^\dagger \ \gamma^{\mu\nu} F_{\mu\nu} -   \varepsilon_1^\dagger \ \gamma_5 \ D\;.
\end{equation}
One then obtains for the bosonic piece
\begin{equation}
  \frac{1}{4}\Tr\  (\mathcal Q \lambda)^\ddagger\  \mathcal Q\lambda = \frac{1}{2}\Tr\ \left( D^2 + \frac{1}{2} \sum_{m,n} (F_{mn})^2 \right),
\end{equation}
whose zero-locus is indeed $D=F_{mn}=0.$

Higgs branch localization requires the addition of an extra $\mathcal{Q}-$exact deformation term
\begin{equation}
\mathcal{L}_{\text{H}}^{\text{def}} = \frac{i}{2}  \mathcal{Q}\ \Tr\  \varepsilon_1^\dagger \gamma_5 \lambda\  H(\phi)\;,
\end{equation}
whose bosonic part is
\begin{equation}
\mathcal{L}_{\text{H}}^{\text{def}}\Big|_{\text{bos}} = -\Tr\ \left(iD + \cos\theta\ F_{23} - \sin\theta\ F_{13}\right)\  H(\phi)\;.
\end{equation}
Upon adding $\mathcal{L}_{\text{YM}}^{\text{def}}$ and $\mathcal{L}_{\text{H}}^{\text{def}},$ the auxiliary field $D$ can be integrated out exactly by performing the Gaussian path integral. Correspondingly, one imposes its field equation $D = i H(\phi).$ The auxiliary field $D$ is thus taken out of its real contour. The bosonic part of the total deformation Lagrangian can then be written as a sum of squares once again:
\begin{align}
\mathcal{L}_{\text{YM}}^{\text{def}}\Big|_{\text{bos}} + \mathcal{L}_{\text{H}}^{\text{def}}\Big|_{\text{bos}} = \frac{1}{2}\Tr\ &\left((F_{12})^2+(F_{14})^2+(F_{24})^2+(F_{34})^2 +\right.\notag \\ &\left. + (-H(\phi) -\sin\theta\ F_{13} + \cos\theta\ F_{23})^2 + (\cos\theta\ F_{13} + \sin\theta\  F_{23})^2 \right)\;,
\end{align}
from which we read off the BPS equations
\begin{equation}\label{BPSeqnYM}
F_{12}=F_{14}=F_{24}=F_{34}=-H(\phi) -\sin\theta\ F_{13} + \cos\theta\ F_{23}=\cos\theta\ F_{13} + \sin\theta\  F_{23}=0\;.
\end{equation}
Note that these equations could have been obtained equivalently from \eqref{generalBPSeqnYM1}-\eqref{generalBPSeqnYM2} by imposing the D-term equation. More explicitly, these equations read in the coordinate frame
\bea\label{explicitBPSeqnYM}
0 &= F_{\varphi\chi} = F_{\varphi\tau} = F_{\chi\tau} \\
F_{\theta \tau} &= \frac{2\eta}{3\xi \ell} F_{\varphi\theta} = -\frac{2\eta}{3\xi \ell} F_{\chi\theta}\\
-\ell^2 H(\phi)&=   \frac{F_{\varphi\theta}}{\sin\theta\ \cos\theta}  = -\frac{F_{\chi\theta}}{\sin\theta\ \cos\theta}  \;.
\eea

Let us next turn our attention to chiral multiplets. We take them to transform under some generic representation $\mathfrak{R}$ of the flavor and gauge group. Let us denote its decomposition in irreducible gauge representations as $\mathfrak{R}=\sum_i \mathcal{R}_i.$ The BPS equations for a single chiral multiplet transforming in representation $\mathcal{R}$ are found by setting to zero each component of the variation of the fermion $\chi$ under the supersymmetry transformation by $\varepsilon_1.$ Subsequently imposing the reality property $\phi^\dagger = \bar\phi$ and $\mathcal{F}^\dagger = \bar{\mathcal{F}}$ and taking appropriate linear combinations, one obtains
\bea \label{BPSeqnmatter}
0 &=(D_4 - D_4^\dagger)\phi \qquad\qquad &
0 &=\cos\theta\ D_2\phi - \sin\theta\ D_1\phi + i D_3\phi\\
0 &=  \mathcal{F}  &
0 &=  3 r \phi + \ell\left( 2 i (\cos\theta\ D_1\phi + \sin\theta\ D_2\phi) - (D_4 + D_4^\dagger)\phi \right) \;,
\eea
Using that
\begin{equation}
D_4 \phi = \left(D_\tau - \frac{\eta}{3\xi\ell}(D_\chi - D_\varphi) + \frac{r}{2\ell} - \frac{i}{3\xi\ell} \fz \right)\phi\;,
\end{equation}
these equations can be written explicitly as (assuming the gauge field is real)
\begin{eqnarray}\label{chiralBPSeqn1}
0 &=& \left(D_\tau - \frac{\eta}{3\xi\ell}(D_\chi - D_\varphi) - \frac{i}{3\xi\ell}\fz\right)\phi\;, \\
0 &=& \mathcal{F} \;, \\\label{chiralBPSeqn2}
0 &=&  r \phi +   i ( D_\varphi\phi +  D_\chi\phi)\;, \\\label{chiralBPSeqn3}
0 &=& \cot\theta\ D_\chi\phi - \tan\theta\ D_\varphi\phi + i D_\theta\phi\;.
\end{eqnarray}

\section{BPS solutions: Coulomb, Higgs and vortices}\label{section_BPSsolutions}
In this section we set out to solve the BPS equations. Let us first recall the Coulomb branch solutions.
\paragraph{Coulomb branch}
The Coulomb branch solution was already mentioned above:
\begin{equation}
D=0\;, \qquad A=\frac{a}{3\xi \ell} d\tau\;.
\end{equation}
As usual $a$ can be taken to lie in the Cartan algebra. Let us verify that there are no solutions to the chiral multiplet equations for positive R-charges. Fourier expanding the chiral field as
\begin{equation}
\phi = \sum_{p,m,n} e^{2\pi i p \tau /3\xi\ell}\ e^{in\varphi}\ e^{im\chi}\ c_{pmn}(\theta)\;,
\end{equation}
one finds from \eqref{chiralBPSeqn1} that only modes for which
\begin{equation}
(a+\fz)\phi = 2\pi p - \eta (m-n)
\end{equation}
can exist. Via a large gauge transformation, we can set $p=0\;.$ Next, equation \eqref{chiralBPSeqn2} further imposes that $r = n+m.$ Finally, equation \eqref{chiralBPSeqn3} reduces to the differential equation
\begin{equation}
\partial_\theta c_{pmn} = -(m\cot\theta - n \tan\theta) c_{pmn}\;,
\end{equation}
which solves to $c_{pmn}(\theta) = \phi_0 (\cos\theta)^{-n} (\sin\theta)^{-m} \;,$ for some constant $\phi_0\;.$ Smoothness at $\theta = 0$ and $\theta = \frac{\pi}{2}$ demands that $m\leq 0$ and $n\leq 0$ respectively. Therefore, for positive R-charges, no solutions exist. For zero R-charge (then $m=n=0$), we find the constant Higgs like solution $\phi=\phi_0,$ if $(a+\fz)\phi =0$.

Next, we study the new solutions which become available upon choosing a non-trivial $H(\phi),$ \ie\  we want to solve \eqref{BPSeqnYM} and \eqref{BPSeqnmatter}. We set the R-charges to zero, $r=0$: the exact non-anomalous R-charge should be restored by giving an imaginary part to the flavor fugacities. We make the standard choice for $H(\phi)$:
\begin{equation}
\label{H(phi)}
H(\phi) = \zeta - \sum_{i,a} T^a_\text{adj} \ \phi_i^\dagger T^a_{\mathcal{R}_i} \phi_i\;,
\end{equation}
where the sum runs over the matter representations $\mathcal{R}_i$ and its generators $T^a_{\mathcal{R}_i.}$ Here, $\zeta$ is adjoint-valued and defined as a real linear combination of the Cartan generators $h_a$ of the Abelian factors in the gauge group
\begin{equation}\label{QexactFI}
\zeta = \sum_{a:\ U(1)} \zeta_a h_a \;.
\end{equation}
We find the following classes of solutions.

\paragraph{Deformed Coulomb branch}
The deformed Coulomb branch is characterized by $\phi=0.$ A solution to the vector multiplet BPS equations \eqref{BPSeqnYM} is then given by
\begin{equation}
F = \zeta \ell^2 \sin\theta\ \cos\theta\ d\theta \wedge \left( d\varphi - d\chi - \frac{2\eta}{3\xi \ell} d\tau  \right)\;,
\end{equation}
which can be integrated to 
\begin{equation}
A = -\zeta \ell^2 \left(\frac{1}{2}\cos^2\theta \left( d\varphi - \eta \frac{d\tau}{3\xi\ell}  \right) + \frac{1}{2}\sin^2\theta \left( d\chi + \eta \frac{d\tau}{3\xi\ell}  \right) \right) + \frac{a}{3\xi\ell}d\tau\;.
\end{equation}

\paragraph{Higgs-like solutions}
Higgs-like solutions are defined by setting $H(\phi)=0.$ Then it follows that also $F_{\mu\nu}=0.$ From above, we know that $\phi=\phi_0$ is a constant constrained by the condition $(a+\fz)\phi_0 = 0\;.$

Solutions to the D-term equations 
\begin{equation}\label{Dtermeqns}
H(\phi)=0\;, \qquad (a+\fz)\phi = 0\;,
\end{equation}
depend both on the gauge group and on the matter representations. Here we will restrict ourselves to cases where the vacuum expectation values of $\phi$ completely break the gauge group.

\paragraph{Vortices}
Each Higgs-like solution is the root of a tower of vortex solutions at the north and south torus. Indeed, using the other BPS equations, the BPS equations \eqref{chiralBPSeqn3} and the last equation in \eqref{explicitBPSeqnYM} become for $\theta\rightarrow 0,$ and introducing $R\equiv \ell \theta,$
\begin{equation}
0 =\left(D_R - \frac{i}{R}D_\chi\right)\phi\;,\qquad H(\phi) = -\frac{1}{R} F_{R\chi}\;,\label{NPvortexeqn}
\end{equation}
which we recognize as the standard (anti)vortex equations on $\mathbb{R}^2.$ Once the solutions to these equations are found, the other BPS equations will complete it to solutions on $\mathbb{R}^2\times T^2.$ The vortex equations cannot be solved analytically, so we shall content ourselves with qualitatively analyzing the behavior of the solutions. We consider the case of a $U(1)$ theory with a single chiral multiplet of (gauge) charge +1. Up to rescalings of the latter, this is the generic case once the gauge group is broken to its maximal torus. Let us start by making the Ansatz
\begin{equation}
\phi = e^{-in\varphi} e^{-im\chi} \phi_0(R)\;, \qquad A=A_\tau(R) d\tau + A_\varphi(R)d\varphi + A_\chi(R) d\chi\;,
\end{equation}
where we didn't include a time dependence since it can be removed by the same large gauge transformation we employed earlier. When $\phi_0 \neq 0$ one finds from \eqref{chiralBPSeqn1} and \eqref{chiralBPSeqn3} the exact relations
\begin{equation}\label{exactrelations}
A_\tau = \frac{1}{3\xi\ell}\Big(\eta \left((A_\chi+m)-(A_\varphi+n) \right) - \fz\Big)\;, \qquad A_\varphi + A_\chi = -(n+m)\;.
\end{equation}
Given these exact relations, all BPS equations are satisfied except for the vortex equations \eqref{NPvortexeqn} themselves:
\begin{equation}
\partial_R \phi_0 - \frac{1}{R}(m+A_\chi)\phi_0 = 0\;, \qquad \zeta - \phi_0^2 = -\frac{1}{R}\partial_R A_\chi\;,
\end{equation}
and moreover it is sufficient to outline the behavior of $A_\chi$ and $\phi.$ When $R\rightarrow 0$ (more precisely, for $R\ll\sqrt{\frac{m}{\zeta}}$), in order to have a smooth connection, one necessarily has $A_\chi\rightarrow 0\;.$ The first equation then further implies that $\phi_0 = B R^m\;.$ In particular we deduce that $m>0.$ From the second equation to leading order in $R$ we deduce that $\partial_R A_\chi = -R\zeta$ and thus $A_\chi = -\frac{\zeta R^2}{2}.$ For $R\rightarrow \infty$ ($R\gg\sqrt{\frac{m}{\zeta}}$), $\phi$ sits in its Higgs vacuum $\phi_0\rightarrow \zeta.$ Then one finds that $A_\chi\rightarrow -m.$ Integrating the field strength over $\mathbb{R}^2,$ one finds that $m$ can be interpreted as the vortex number $\frac{1}{2\pi}\int F = -m. $  When approximating $R^{-1}F_{R\chi}$ by a step function of height $-\zeta,$ we immediately find a measure for the size of the vortex to be $\sqrt{\frac{m}{\zeta}}$. For sufficiently large values of $\zeta$ the vortex shrinks to zero size and the first order approximations we took are justified. Momentarily, we will give an interpretation to $n$ as well.

It is noteworthy that $A_\tau$ only asymptotically sits in its Higgs vacuum: for $R\rightarrow 0$ one finds $A_\tau = \frac{1}{3\xi\ell}(2m\eta - \fz) - \frac{\eta}{3\xi\ell}\zeta R^2\;.$

One can similarly analyze the behavior for $\theta\rightarrow \frac{\pi}{2}.$ Introducing $\rho = \ell \left(\frac{\pi}{2}-\theta \right),$ one again finds the vortex equations among the BPS equations
\begin{equation}\label{SPvortexeqn}
0 = \left(D_\rho - \frac{i}{\rho}D_\varphi\right)\phi\;,\qquad H(\phi) =- \frac{1}{\rho} F_{\rho\varphi}\;.
\end{equation}
Let us also here analyze the qualitative behavior for the case of a $U(1)$ theory with a single chiral of charge +1. Starting by making the Ansatz
\begin{equation}
\phi = e^{-in\varphi} e^{-im\chi} \phi_0(\rho)\;, \qquad A=A_\tau(\rho) d\tau + A_\varphi(\rho)d\varphi + A_\chi(\rho) d\chi\;.
\end{equation}
we rediscover the exact relations \eqref{exactrelations} which solve all BPS equations but
\begin{equation}
\partial_\rho \phi_0 - \frac{1}{\rho}(n+A_\varphi)\phi_0 = 0\;, \qquad \zeta - \phi_0^2 = -\frac{1}{\rho}\partial_\rho A_\varphi\;.
\end{equation}
For $\rho\ll \sqrt{\frac{n}{\zeta}}\;,$ smoothness demands that $A_\varphi\rightarrow 0\;.$ In this region we then find from the first equation that $\phi_0 = B^\prime \rho^n,$ implying that $n>0.$ To leading order in $\rho$ the second equations teaches that $\partial_\rho A_\varphi = -\rho\zeta$ and thus $A_\varphi = -\frac{\zeta \rho^2}{2}.$ For $\rho\gg\sqrt{\frac{n}{\zeta}}\;,$ we have $\phi_0\rightarrow \zeta$ and $A_\varphi\rightarrow -n.$ Since integrating over $\mathbb{R}^2$ gives $\frac{1}{2\pi}\int F = -n,$ we interpret $n$ as the vortex number at the south torus.

Also here $A_\tau$ sits only asymptotically in its Higgs vacuum. Note also that in the intermediate region both solutions glue together appropriately.

For smaller values of $\zeta$ both the presence of curvature in and the finite volume of space will start affecting the solutions. However, we can derive an exact bound by integrating $H(\phi)$ over spacetime and using the last BPS equation in \eqref{explicitBPSeqnYM}
\begin{eqnarray}
\zeta \vol(S^3\times S^1) &\geq& \int_{S^3\times S^1} H(\phi)\  d\vol(S^3\times S^1) \notag \\
 &=& 4\pi^2\ell \vol(S^1) \int_0^{\frac{\pi}{2}} d\theta \partial_\theta A_\varphi=-4\pi^2\ell \vol(S^1) \int_0^{\frac{\pi}{2}} d\theta \partial_\theta A_\chi
\end{eqnarray}
Here we used that on vortex solutions $0\leq H(\phi)\leq \zeta$ and that vortex solutions don't have $\theta$ dependence. Defining the vorticities as the winding numbers of $\phi$ around $\chi,\varphi$ respectively and employing the analysis at the core of the vortex, we then find that 
\begin{equation}\label{bound}
4\pi^2\ell (n+m)\leq \zeta \vol(S^3) \Rightarrow n+m\leq \zeta \frac{\ell^2}{2}\;.
\end{equation}
One observes that for finite values of $\zeta$ only a finite number of vortices are supported on $S^3\times S^1.$ The bound is saturated precisely when $\phi$ vanishes; the solution is then described by the deformed Coulomb branch solution. 

We thus find essentially the same interpretation as in \cite{Benini:2013yva}. Upon increasing $\zeta$ from 0 to $+\infty$ the original Coulomb branch solution is deformed into the deformed Coulomb branch and each time $\zeta$ crosses a bound \eqref{bound}, a collection of new vortices branches out.

\section{Computation of the index}\label{section_computationindex}
In the previous section, we found various classes of BPS solutions. The final steps in the computation of the index using localization, are then to first evaluate the classical action on and the one-loop determinants of quadratic fluctuations around these solutions, and next integrate and/or sum over the space of BPS configurations.

\subsection{One-loop determinants from an index theorem}
Although the computation of the one-loop determinants can be straightforwardly performed on the Coulomb branch (in a Lagrangian theory like the ones at hand) by enumerating letters, constructing the single letter partition function, subsequently plethystically exponentiating these and finally imposing the Gauss law constraint by projecting onto gauge singlets, the computation on non-constant configurations is most easily performed using an equivariant index theorem for transversally elliptic operators \cite{Atiyah:1974}. The idea is to bring the problem in cohomological form, and make use of the fact that, via the equivariant index theorem, only the fixed points of the equivariant spatial rotations contribute to the one-loop determinants. A detailed discussion can be found in \cite{Pestun:2007rz}.

Recall from \eqref{Qsquared} that the supercharge squares to
\begin{eqnarray}\label{Qsquared2}
\delta^2_{\varepsilon_1}&=& -\frac{2}{3\xi\ell}\left[ -3\xi\ell\ \mathcal{L}^A_{\partial_{\hat\tau}} + 6\xi i\ \mathcal{L}^A_{\frac{1}{2}\left(\partial_{\hat\varphi} + \partial_{\hat\chi}\right)} + 2\eta\ \mathcal{L}^A_{\frac{1}{2}\left(\partial_{\hat\chi} - \partial_{\hat\varphi}\right)}  + 3\xi R + i\sum_j \fz_j F_j \right]\\
&=& -\frac{2}{3\xi\ell}\left[ -3\xi\ell\ \mathcal{L}^A_{\partial_{\hat\tau}} + (3\xi i - \eta)\  \mathcal{L}^A_{\partial_{\hat\varphi}} + (3\xi i + \eta)\  \mathcal{L}^A_{\partial_{\hat\chi}}  + 3\xi R + i\sum_j \fz_j F_j \right]\;.
\end{eqnarray}
where we used the value $\beta=3\xi.$ An important observation is that $\delta^2_{\varepsilon_1}$ precisely equals (upon properly identifying the equivariant parameters\footnote{The precise identifications between the equivariant parameters here and those on the squashed three-sphere (see for example expression C.3 in \cite{Benini:2013yva}) are $b=3\xi i -\eta,$ $b^{-1}=3\xi i +\eta$ up to a constant rescaling.}) the square of the supercharge used in the localization on $S^3_b$ in \cite{Drukker:2012sr} (see also \cite{Benini:2013yva}) with an additional free motion along the temporal circle generated by  $ -3\xi\ell\ \mathcal{L}^A_{\partial_{\hat\tau}}.$ Thus, taking into account the Kaluza-Klein modes along the temporal circle, the computation of the equivariant index on $S^3\times S^1$ can be effectively reduced to that on a squashed three-sphere. This latter computation was performed in \cite{Drukker:2012sr} (see also appendix C of\cite{Benini:2013yva}) and involves a further reduction along the Hopf fiber. The base space of the double reduction, which is topologically a two-sphere, has two fixed points (one at $\theta=0$ which we call North and one at $\theta=\frac{\pi}{2}$ (South)) under the reduction of the spatial rotations appearing in $\delta^2_{\varepsilon_1}.$  The equivariant index only receives contributions from these two points.

Introducing the equivariant parameter for gauge transformations
\begin{equation}
i\hat{a} = -3\xi\ell(-iA_\tau) + 3\xi i (-i(A_\varphi+A_\chi)) + \eta(-i(A_\chi-A_\varphi))\;,
\end{equation}
we can now immediately write the one-loop determinant for the vector multiplet
\begin{equation}
Z_{\text{1-loop}}^{\text{vector}}\  \text{``}=\text{''}\  \prod_{\substack{n,m\in\mathbb{Z}\\ \alpha \in\fg}} \left( \pi i n -\frac{i}{2}(3\xi i -\eta) m - \frac{i}{2} \alpha(\hat{a}_N) \right)^{1/2} \left( \pi i n -\frac{i}{2}(3\xi i +\eta) m - \frac{i}{2} \alpha(\hat{a}_S) \right)^{1/2}\;,
\end{equation}
where $\alpha\in\mathfrak{g}$ denotes the roots of the gauge algebra $\mathfrak{g}$. Compared to the unregularized vector multiplet one-loop determinant on the squashed three-sphere an extra  product over the integer $n$ appears, which precisely captures the contribution of the Kaluza-Klein modes along the temporal circle. Regularizing the infinite products results in
\begin{eqnarray}
Z_{\text{1-loop}}^{\text{vector}} &=& \left[\left( t^3 y^{-1}\; ;\; t^3 y^{-1} \right)_\infty \left( t^3 y\; ;\; t^3 y \right)_\infty\right]^{\rank \fg} \prod_{\alpha \neq 0} \left(1-e^{i\alpha(\hat a_N)}\right)^{1/2} \left(1-e^{i\alpha(\hat a_S)}\right)^{1/2} \notag \\&&\times\prod_{\alpha \neq 0} \left( t^3 y^{-1}\  e^{ i\alpha(\hat{a}_N)}\; ;\; t^3 y^{-1} \right)_\infty \left( t^3 y\  e^{ i\alpha(\hat{a}_S)}\; ;\; t^3 y \right)_\infty \;,
\end{eqnarray}
in terms of the infinite q-Pochhammer symbol $(z,q)_\infty = \prod_{j=0}^{\infty} (1-z q^j).$ Using the standard plethystic exponential, it can be written as
\begin{multline}
Z_{\text{1-loop}}^{\text{vector}} = \prod_{\alpha \neq 0} \left(1-e^{i\alpha(\hat a_N)}\right)^{1/2} \left(1-e^{i\alpha(\hat a_S)}\right)^{1/2}\\
\times\text{P.E.}\left[ -\frac{t^3y^{-1}}{1-t^3y^{-1}} \left(\rank\fg + \sum_{\alpha\neq 0}e^{i\alpha(\hat{a}_N)} \right) -\frac{t^3y}{1-t^3y} \left(\rank\fg + \sum_{\alpha\neq 0}e^{i\alpha(\hat{a}_S)} \right)\right]\;,
\end{multline}
For all BPS configurations we will consider $\hat{a}_N=\hat{a}_S=\hat a.$ The vector multiplet one-loop determinant simplifies then further to
\begin{equation}\label{oneloop_vector1}
Z_{\text{1-loop}}^{\text{vector}} =\prod_{\alpha \neq 0} \left(1-e^{i\alpha(\hat a)}\right)\  \text{P.E.}\left[ -\left(\frac{t^3y^{-1}}{1-t^3y^{-1}}  +\frac{t^3y}{1-t^3y} \right)\left(\rank\fg + \sum_{\alpha\neq 0}e^{i\alpha(\hat{a})} \right)\right]  \;.
\end{equation}
Observing that $-\left(\frac{t^3y^{-1}}{1-t^3y^{-1}}  +\frac{t^3y}{1-t^3y} \right) = \frac{2t^6 - t^3(y+y^{-1})}{(1-t^3y^{-1})(1-t^3y)},$ one recognizes the single letter partition function of the vector multiplet \cite{Romelsberger:2007ec}. Using that $-\left(\frac{t^3y^{-1}}{1-t^3y^{-1}}  +\frac{t^3y}{1-t^3y} \right)=1- \frac{1-t^6}{(1-t^3y^{-1})(1-t^3y)}$ it can be written alternatively as \cite{Dolan:2008qi}
\begin{eqnarray}\label{oneloop_vector2}
Z_{\text{1-loop}}^{\text{vector}}
&=& \left((t^3y\ ;\ t^3y)_\infty (t^3y^{-1}\ ;\ t^3y^{-1})_\infty \right)^{\rank \fg} \prod_{\substack{ n,m\geq 0 \\ \alpha\neq 0}} \frac{1-e^{i\alpha(\hat{a})} (t^3y)^n (t^3y^{-1})^m}{(1-e^{i\alpha(\hat{a})} (t^3y)^{n+1} (t^3y^{-1})^{m+1})}  \notag  \\
&=& \left((t^3y\ ;\ t^3y)_\infty (t^3y^{-1}\ ;\ t^3y^{-1})_\infty \right)^{\rank \fg} \prod_{\alpha\neq 0} \frac{1}{\Gamma(e^{i\alpha(\hat a)},t^3y,t^3y^{-1})}\; ,
\end{eqnarray}
in terms of the standard elliptic gamma function 
\begin{equation}
\Gamma(z, p,q) = \prod_{j,k\geq 0}\frac{1-p^{j+1}q^{k+1}/z}{1-p^jq^k z}\;.
\end{equation}

For the one-loop determinant of a chiral multiplet of R-charge $r$ transforming in gauge representation $\mathcal{R}$ we find the unregularized expression
\begin{equation}
Z_{\text{1-loop}}^{\text{chiral}} \text{``}= \text{''} \prod_{w \in \mathcal{R}} \prod_{\substack{n,m\in\mathbb{Z}\\p\geq 0}} \frac{-\pi i n + \frac{i}{2}(3\xi i +\eta) m + \frac{i}{2}(3\xi i -\eta)(p+1) + \frac{3}{2}\xi r + \frac{i}{2}w(\hat{a}_S) + \frac{i}{2} \fz }{-\pi i n + \frac{i}{2}(3\xi i -\eta) m - \frac{i}{2}(3\xi i +\eta)p + \frac{3}{2}\xi r + \frac{i}{2}w(\hat{a}_N) + \frac{i}{2} \fz} \;,
\end{equation}
where $w\in\mathcal{R}$ denotes the weights of the representation $\mathcal{R}$. Also here the extra contribution of the Kaluza-Klein modes along the temporal circle is given by the infinite product over the integer $n.$

When $\hat{a}_N=\hat{a}_S=\hat a,$ it can be regularized to
\begin{equation}\label{oneloop_chiral}
Z_{\text{1-loop}}^{\text{chiral}}= \prod_{w \in \mathcal{R}} \Gamma\left( t^{3r}e^{ - i w(\hat{a}) - i \fz }  \ , \ t^3 y\ , \ t^3y^{-1} \right) = \prod_{w\in\mathcal{R}}\text{P.E.}\left[ \frac{ t^{3r} e^{-i\fz}\ e^{-iw(\hat a)} - t^{3(2-r)}e^{i\fz}\ e^{iw(\hat a)} }{(1-t^3y^{-1})(1-t^3y)}\right] \;,
\end{equation} 
where again one recognizes the correct single letter partition function \cite{Romelsberger:2007ec}. 

\subsection{Coulomb branch}
Let us first briefly recall the Coulomb branch expression \cite{Romelsberger:2007ec,Dolan:2008qi}. As was mentioned before, both the gauge and matter Lagrangians are $\mathcal{Q}$-exact, and we only have to evaluate the Fayet-Iliopoulos term:
\begin{equation}
S_{FI} = \frac{-i\ell}{2 \vol(S^3)}\Tr_{FI} \int_{S^3\times S^1} \left(D+\frac{2}{\ell} A_\tau\right) \dvol(S^3\times S^1) = -i\Tr_{FI} a\;.
\end{equation}
The equivariant parameter for the gauge transformation $i\hat{a} = 3\xi i \ell A_\tau + 3\xi (A_\varphi+A_\chi) - i \eta(A_\chi-A_\varphi)$ simply gives $\hat{a}_N=\hat{a}_S=a.$ The one-loop determinants \eqref{oneloop_vector2} and \eqref{oneloop_chiral} are thus
\begin{equation}\label{oneloop_Coulomb}
Z_{\text{1-loop}}^{\text{vector}} = \frac{\left((t^3y\ ;\ t^3y)_\infty (t^3y^{-1}\ ;\ t^3y^{-1})_\infty \right)^{\rank \fg}}{ \prod_{\alpha\neq 0}\Gamma(e^{i\alpha(a)},t^3y,t^3y^{-1})}\; , \qquad Z_{\text{1-loop}}^{\text{chiral}}= \prod_{w \in \mathcal{R}} \Gamma\left( t^{3r}e^{ - i w(a) - i \fz }  \ , \ t^3 y\ , \ t^3y^{-1} \right) \;,
\end{equation}
and the index can be computed by
\begin{equation}\label{matrixintegral_Coulomb}
I = \frac{1}{|\mathcal{W}|} \oint \left(\prod_{j=1}^{\rank G}\frac{dz_j}{2\pi i z_j}\right) \ e^{ i\Tr_{FI} a}\  Z_{\text{1-loop}}\;,
\end{equation}
where $|\mathcal{W}|$ denotes the dimension of the Weyl group of the gauge group $G,$ $z_j = e^{ia_j}$ and the integration contour is along the unit circle. Note that the quantized nature of the FI parameter can now be seen to ensure that the integrand remains meromorphic. We should also mention that the usual Vandermonde determinant cancels against the contribution of the gauge-fixing ghosts.

\subsection{Deformed Coulomb branch}
Next, we study the situation on the deformed Coulomb branch. Using $D=iH(\phi)$ which equals $i\zeta$ on this solution, we obtain for the classical action
\begin{equation}
S_{FI} =  -i\Tr_{FI} \left(a + i \frac{3}{2}\xi \ell^2 \zeta\right)\;,
\end{equation}
where we also used that $A_\tau = \frac{1}{3\xi\ell} \left( a +\frac{\eta \ell^2}{2} \zeta \cos 2\theta \right).$
For the equivariant parameter $\hat{a}$ we find
\begin{equation}
\hat a = \hat a_N = \hat a_S = a + i \frac{3}{2}\xi \ell^2 \zeta\;.
\end{equation}
Both in the classical action and the one-loop determinants, the effect of the deformation is seen to be given by an imaginary shift of the holonomy variable $a\rightarrow a + i \frac{3}{2}\xi \ell^2 \zeta$ or thus $z=e^{ia}\rightarrow z \ t^{\frac{3}{2}\ell^2\zeta}.$ When used in the matrix integral \eqref{matrixintegral_Coulomb}, one effectively changes the radius of the integration contour. Indeed, since $t<1,$ one finds that the contour shrinks (grows) for $\zeta\rightarrow +\infty$ ($\zeta\rightarrow -\infty$). When turning on the deformation parameter $\zeta,$ the integral remains constant as long as no poles of the integrand are crossed. Moreover, one can understand by looking at the bound \eqref{bound}, that the jumps in the integral, which are equal to the residues of the crossed poles, are precisely the contributions of the newly available vortex configurations. We thus recover the same picture as was found in \cite{Benini:2013yva} in three dimensions.

Of particular interest is the situation where the index is expressed only in terms of vortices. This can be achieved if there exists a certain limit for the parameters $\zeta^a\rightarrow \pm \infty$ such that the deformed Coulomb branch is suppressed. In view of the shrinking/growing contour, such suppression can be obtained heuristically if the residue at the origin or infinity vanishes.

\subsection{Higgs branch and vortices}
For finite values of the deformation parameters $\zeta^a,$ the deformed Coulomb branch contribution of the previous subsection is complemented by finite size vortex configurations satisfying the bound \eqref{bound}. Evaluation of their classical action can be done exactly in a gauge $A_\theta=0,$ using the BPS equation \eqref{explicitBPSeqnYM}, the behavior of the vortices in their core and the exact relations \eqref{exactrelations}. We then find
\begin{equation}\label{resclassical}
S_{FI} = -i\Tr_{FI} \left(3\xi i (n+m) -\fz + \eta(m-n) \right)\;,
\end{equation}
where the vortex numbers $m,n$ are GNO quantized elements of the coweight lattice. For the evaluation of the one-loop determinants, we first consider the contribution from the off-diagonal W-bosons and those chiral multiplets that do not acquire a vacuum expectation value. Their one-loop determinant is simply found by inserting the equivariant parameter evaluated on the vortex background
\begin{equation}
\hat a = \hat a_N = \hat a_S = -\fz + 3\xi i(m+n) + \eta (m-n)\;,
\end{equation}
in the expressions for the one-loop determinants \eqref{oneloop_vector2}, with the contribution of the diagonal vector multiplets, \ie\  $\left((t^3y\ ;\ t^3y)_\infty (t^3y^{-1}\ ;\ t^3y^{-1})_\infty \right)^{\rank \fg},$ removed, and \eqref{oneloop_chiral}. The $\rank \mathfrak{g}$ chiral multiplets that do get a VEV are eaten by the diagonal vector multiplets, which in turn become massive, via the Higgs mechanism. As was explained in \cite{Doroud:2012xw}, the one-loop determinant of this paired system is found as the residue of the product of their one-loop determinants. In total one thus finds
\begin{equation}\label{resvectoroneloop}
Z_{\text{1-loop}}^{\text{vector}} = \frac{1}{ \prod_{\alpha\neq 0}\Gamma\left(e^{i\alpha(a_H)}\  \left(t^3y\right)^{ \alpha(m)}\   \left(t^3y^{-1}\right)^{\alpha(n)},t^3y,t^3y^{-1}\right)}\; ,
\end{equation}
and 
\begin{multline}\label{reschiraloneloop}
Z_{\text{1-loop}}^{\text{chiral}}= \left((t^3y\ ;\ t^3y)_\infty (t^3y^{-1}\ ;\ t^3y^{-1})_\infty \right)^{\rank \fg} \\ \times \Res_{a\rightarrow a_H}\prod_{w \in \mathcal{R}} \Gamma\left(t^{3r} e^{ - i w(a)- i \fz }\  \left(t^3y\right)^{-w(m)} \ \left(t^3y^{-1}\right)^{-w(n)}  \ , \ t^3 y\ , \ t^3y^{-1} \right) \;,
\end{multline}
where $a_H$ is the holonomy evaluated in its Higgs vacuum. 

It is clear from \eqref{resclassical}, \eqref{resvectoroneloop} and in particular \eqref{reschiraloneloop}, that when adding the contribution of the vortices satisfying the bound \eqref{bound} to the deformed Coulomb branch integral, we precisely recover the original Coulomb branch expression, since they precisely contribute the residues of the crossed poles. Since the deformation parameters enter our analysis via a $\mathcal{Q}$-exact piece, such picture was expected.

\paragraph{Elliptic vortex partition function} Let us now send the deformation parameters $\zeta^a$ to infinity in such a way that the contribution of the deformed Coulomb branch vanishes. The index is then described purely in terms of point-like vortices which wrap the torus and have arbitrary vortex numbers. The elliptic uplift of the standard vortex partition function \cite{Shadchin:2006yz} describes their total contribution and can be independently computed by considering the theory on $\mathbb{R}^2_\epsilon \times T^2_\tau$ in the $\Omega$-background. The plane $\mathbb{R}^2$ is effectively compactified, since it is rotated as we go around either cycle of the torus. The resulting elliptic vortex partition function $Z_{\text{vortex}}$ can depend on the rotational parameter $\epsilon,$ the complex structure of the torus $\tau,$ flavor fugacities $g$ and a fugacity coupling to leftmoving fermion number. This is all the two dimensional analog of the elliptic instanton partition function obtained by studying the theory on $\mathbb{R}_{\epsilon_1,\epsilon_2}^4\times T^2_\tau,$ see for example \cite{Hollowood:2003cv}.

In the computation of the partition function in this limit, there are three contributions to be considered. First, there is the classical action evaluated on the vortex configuration \eqref{resclassical} which splits into an overall classical action 
\begin{equation}
S_{FI} = -i\Tr_{FI} \left( a_H \right)\;,
\end{equation}
and a weighting factor for the vortex partition functions
\begin{equation}
e^{-S_{\text{v}}} = (t^3y)^{\Tr_{FI} m } , \qquad e^{-S_{\text{av}}} = (t^3y^{-1})^{\Tr_{FI} n }\;.
\end{equation}
Second, the contribution of the off-diagonal vectormultiplets and the chiral multiplets not taking on a vacuum expectation value is as in the Coulomb branch \eqref{oneloop_Coulomb}, but evaluated on the Higgs branch location, \ie\ $a\rightarrow a_H,$ and with the contribution of the diagonal vector multiplets removed. The contributions of the $\rank \mathfrak{g}$ chiral multiplets acquiring a vacuum expectation value and the diagonal vector multiplets cancel each other. Third, there is the vortex partition function itself. Its parameters can be read off from \eqref{Qsquared2}:
\begin{eqnarray}
\epsilon_N = 3\xi i + \eta\;, \;\; &\tau_N = \frac{3\xi i -\eta}{2\pi} + i (-i)\;,\;\; &g_N = a_H + \sum_j\fz_j F_j\;,\\
\epsilon_S = 3\xi i - \eta\;, \;\; &\tau_S = \frac{3\xi i +\eta}{2\pi}+i(-i)\;,\;\; &g_S = a_H + \sum_j\fz_j F_j\;.
\end{eqnarray}
The extra factor of $i$ in the modular parameter is explained by the fact that in our setup $\Delta\sim \partial_\tau$ while the momenta are $P_{\varphi},P_{\chi}\sim i \partial_{\varphi}, i\partial_{\chi}.$ The final expression for the index as obtained by Higgs branch localization is thus
\begin{align}\label{finalresult}
I =  \sum_{\text{Higgs vacua}}  e^{i\Tr_{FI} \left( a_H \right)}\  Z_{\text{1-loop}}^\prime\  Z_{\text{v}} \ Z_{\text{av}}\;,
\end{align}
where the sum runs over solutions to the D-term equations \eqref{Dtermeqns} and the one-loop determinant excludes the chiral multiplets acquiring a VEV and the diagonal vector multiplets. Finally,
\begin{eqnarray}
Z_{\text{v}} &=& Z_{\text{vortex}}\left((t^3y)^{\Tr_{FI} \cdot }\ ;\ t^3y, t^3y^{-1}, e^{i (a_H+\sum_j\fz_j F_j)} \right)\\
Z_{\text{av}} &=& Z_{\text{vortex}}\left((t^3y^{-1})^{\Tr_{FI} \cdot }\ ;\ t^3y^{-1}, t^3y, e^{i (a_H+\sum_j\fz_j F_j)} \right)\;.
\end{eqnarray}
Here the first argument encodes the weight of the vortex sum and is given as an exponentiated linear function on the gauge algebra, the second and third argument are the exponentiated rotational parameter, $e^{i\epsilon},$ and complex structure, $q=e^{2\pi i \tau},$ respectively, and the last argument is the exponentiated flavor equivariant parameter.

\section{Matching the Coulomb branch expression}\label{section_mathcingCB}
In this section we give some examples of how manipulating the Coulomb branch integral gives rise to our Higgs branch result \eqref{finalresult}.

\paragraph{Free chiral multiplet}
For completeness, let us first mention the factorization of the simplest theory, namely the free chiral. Its index was given in \eqref{oneloop_chiral} and can be factorized as \cite{Razamat:2013opa}
\begin{eqnarray}
I=\Gamma\left( t^{3r} \zeta  \ , \ t^3 y\ , \ t^3y^{-1} \right) &=& \Gamma\left( t^{3r} \zeta, t^3y^{-1}, t^6  \right) \Gamma\left( t^{3r+3} y \zeta, t^3y, t^6  \right) \\
& =& \Gamma\left( t^{3r+3} y^{-1}\zeta, t^3y^{-1}, t^6  \right) \Gamma\left( t^{3r} \zeta, t^3y, t^6  \right) \;. 
\end{eqnarray}

\paragraph{U(1) gauge theory}
Next, we consider the example of a $U(1)$ gauge theory with an equal number $N$ of fundamental and antifundamental chiral multiplets, which is necessary to cancel the $U(1)_{\text{gauge}}$ $U(1)_{\text{gauge}}$ $U(1)_{\text{gauge}}$ anomaly. The $U(1)_{R}$ $U(1)_{R}$ $U(1)_{\text{gauge}}$ anomaly then also cancels. The non-anomalous R-charge assignment is determined by requiring the $U(1)_{R}$ $U(1)_{\text{gauge}}$ $U(1)_{\text{gauge}}$ anomaly to vanish. This anomaly is obviously proportional to the R-charge of the chiral fermion, namely $r-1,$ which implies that one should take $r=1.$ Note that these are not the superconformal R-charges of the free IR theory, which equal $r=\frac{2}{3}.$

The matrix integral \eqref{matrixintegral_Coulomb} reads explicitly
\begin{equation}
I=(p,p)_\infty\ (q,q)_\infty \oint \frac{dz}{2\pi i z} z^{\xi_{FI}}\prod_{\alpha=1}^{N} \Gamma(z^{-1}\zeta_\alpha (pq)^{r/2},p,q) \ \prod_{\beta=1}^{N} \Gamma(z\tilde\zeta_\beta^{-1} (pq)^{r/2}, p,q)\;,
\end{equation}
where we introduced the notation that $p=t^3 y$ and $q=t^3y^{-1}.$ We introduced fugacities $\zeta_\alpha$ and $\tilde \zeta_\beta$ for the $SU(N)\times SU(N)$ flavor symmetry. For notational simplicity, let us absorb the R-charges in the flavor fugacities as $Z_\alpha = \zeta_\alpha (pq)^{r/2}$ and  $\tilde Z_\beta^{-1} = \tilde\zeta_\beta^{-1} (pq)^{r/2}.$

The fundamentals contribute zeros at $z=p^{-\kappa-1} q^{-\lambda-1} Z_\gamma $ and poles at $z=p^\kappa q^\lambda Z_\gamma\;.$ The antifundamentals have zeros at $z=p^{\kappa+1} q^{\lambda+1} \tilde Z_\delta$ and poles at $z= p^{-\kappa} q^{-\lambda}\tilde Z_\delta.$ Picking up the poles inside the unit circle\footnote{Here and in the next examples we are not careful about the pole at the origin. If it has a non-zero residue, it would give rise to a not completely suppressed deformed Coulomb branch contribution.}, \ie \ the poles arising from the fundamentals, we obtain using the formulas in appendix \ref{section_appendixElliptic}
\begin{multline}
I = \sum_{\gamma=1}^{N} Z_\gamma^{\xi_{FI}}\prod_{\substack{\alpha=1\\ \alpha\neq \gamma}}^{ N} \Gamma( Z_\gamma^{-1} Z_\alpha,p,q) \ \prod_{\beta=1}^{N} \Gamma( Z_\gamma\tilde Z_\beta^{-1},p,q) \\
\times \sum_{\kappa,\lambda\geq 0} (p^\kappa q^\lambda)^{\xi_{FI}} \left(pq\right)^{\kappa\lambda  N} \prod_{\alpha=1}^N(\tilde Z^{-1}Z_\alpha)^{-\kappa\lambda}\ \frac{\prod_{\beta=1}^{ N} \prod_{j=0}^{\lambda-1} \theta(q^{j}  Z_\gamma\tilde Z_\beta^{-1} , p)\ \prod_{i=0}^{\kappa-1}\theta(p^{i}  Z_\gamma\tilde Z_\beta^{-1} ,q) }{ \prod_{\alpha=1}^{N}  \prod_{j=1}^{\lambda} \theta(q^{-j} Z_\gamma^{-1} Z_\alpha , p)\ \prod_{i=1}^{\kappa}\theta(p^{-i}  Z_\gamma^{-1} Z_\alpha ,q)}\;.
\end{multline}
The intertwining factor vanishes as expected when reinstating the non-anomalous R-charges,
\begin{equation}
\left(pq\right)^{N} \prod_{\alpha=1}^N(\tilde Z^{-1}Z_\alpha)^{-1} = \left(\left(pq\right)^{1-r}\right)^{N} = 1,
\end{equation} 
where we used that $\prod_\alpha \zeta_\alpha = \prod_\alpha \tilde\zeta_\alpha=1.$ We then find
\begin{equation}
I = \sum_\gamma Z_{\text{cl}}^{(\gamma)}\  Z_{\text{1-loop}}^{\prime (\gamma)} \ Z_{\text{v}}^{(\gamma)}\  Z_{\text{av}}^{(\gamma)},
\end{equation}
where the classical and one-loop contribution are given by 
\begin{eqnarray}
Z_{\text{cl}}^{(\gamma)} &=&  Z_\gamma^{\xi_{FI}} \\
Z_{\text{1-loop}}^{\prime (\gamma)} &=& \prod_{\substack{\alpha=1\\ \alpha\neq \gamma}}^{ N} \Gamma( Z_\gamma^{-1} Z_\alpha,p,q) \ \prod_{\beta=1}^{N} \Gamma( Z_\gamma\tilde Z_\beta^{-1},p,q)\;.
\end{eqnarray}
The vortex contributions can be written as
\begin{equation}
Z_{\text{v}}^{(\gamma)} = Z_{\text{vortex}}^{(\gamma)}\left( p^{\xi_{FI}} \ ; \ p,q, \zeta_\alpha, \tilde \zeta_\beta  \right)\;, \qquad Z_{\text{av}}^{(\gamma)} = Z_{\text{vortex}}^{(\gamma)}\left( q^{\xi_{FI}} \ ; \ q,p, \zeta_\alpha, \tilde \zeta_\beta  \right)\;,
\end{equation}
in terms of the vortex membrane partition function
\begin{equation}
Z_{\text{vortex}}^{(\gamma)}\left( L\ ; \ e^{i\epsilon},q=e^{2\pi i \tau}, a_\alpha, b_\beta  \right) = \sum_{\kappa\geq 0} L^\kappa \frac{ \prod_{j=0}^{\kappa-1}\prod_{\beta=1}^{ N}\theta((e^{i\epsilon})^{j} \  A_\gamma B_\beta^{-1} ,q) }{ \prod_{j=1}^{\kappa}\theta((e^{i\epsilon})^{-j} ,q) \prod_{\substack{\alpha=1\\\alpha\neq \gamma}}^{N}  \theta((e^{i\epsilon})^{-j}\   A_\gamma^{-1}\  A_\alpha ,q)}\;,
\end{equation}
where $A_\alpha = a_\alpha \left( e^{i\epsilon} q \right)^{\frac{1}{2}}$ and $B_\beta = b_\beta \left( e^{i\epsilon} q \right)^{-\frac{1}{2}}.$

\paragraph{U(N) gauge theory}
For a $U(N_c)=U(1)\times SU(N_c)$ gauge theory with $N_f=N_a=N$ fundamentals and antifundamentals, we should check cancellation of two potential anomalies, namely the $U(1)_{\text{gauge}}$ $U(1)_{\text{gauge}}$ $U(1)_{R}$ anomaly and the $SU(N_c)$ $SU(N_c)$ $U(1)_R$ anomaly. The $U(1)_{R}$ $U(1)_{R}$ $U(1)_{\text{gauge}}$ anomaly cancels thanks to $N_f=N_a.$ While the first anomaly is again proportional to $r-1,$ and thus imposes that $r=1,$ the second one leads to the usual R-charge assignment $r=\frac{N_f-N_c}{N_f}.$ These are not compatible for $N_c \neq 0.$ One should thus not hope to achieve factorization in a $U(N_c)$ theory with only fundamentals and antifundamentals. One resolution, also used in two dimensions \cite{Gadde:2013dda,Benini:2013xpa}, might be to add extra matter to cancel the anomaly. We will not pursue this resolution here.

\paragraph{Associated Cartan theory}\label{section_abelianization}
At first sight, Higgs branch localization breaks down in the absence of an abelian factor in the gauge group since one cannot introduce the Fayet-Iliopoulos parameter $\zeta$ of \eqref{QexactFI}, which played such an essential role. However, we will now argue that one can associate to any theory with gauge group $G$ a theory with gauge group $U(1)^{\rank \mathfrak{g}}$ with equal index up to numerical and other holonomy independent factors. A similar observation was made in \cite{Halverson:2013eua} for the two-sphere partition function. This associated Cartan theory can be subjected to Higgs branch localization.

First, one remarks that the integration measure of the matrix integral \eqref{matrixintegral_Coulomb} for gauge group $G$ is naturally equal to that of $U(1)^{\rank \mathfrak{g}}$ up to the numerical prefactor $|\mathcal{W}|^{-1}.$ Next, the one-loop determinant of a chiral field in gauge representation $\mathcal{R}$ of $G$ can be equivalently thought of as the product of one-loop determinants of chiral fields with $U(1)^{\rank \mathfrak{g}}$ charges determined by the weights $w\in \mathcal{R}.$ Finally, using the simple observation that the one-loop determinant of the vector multiplet can be rewritten as
\begin{eqnarray}
Z_{\text{1-loop}}^{\text{vector}}
&=& \frac{\left((t^3y\ ;\ t^3y)_\infty (t^3y^{-1}\ ;\ t^3y^{-1})_\infty \right)^{\rank \fg}}{ \prod_{\alpha\neq 0} \Gamma(e^{i\alpha(\hat a)},t^3y,t^3y^{-1})}\\
&=& \left((t^3y\ ;\ t^3y)_\infty (t^3y^{-1}\ ;\ t^3y^{-1})_\infty \right)^{\rank \fg} \prod_{\alpha\neq 0} \Gamma(t^6 e^{-i\alpha(\hat a)},t^3y,t^3y^{-1})\; ,
\end{eqnarray}
where we used the elliptic gamma function identity $\Gamma(z,p,q) \ \Gamma(pq/z, p,q)=1,$ one can equivalently think of the vector one-loop determinant (up to a holonomy independent prefactor) as the product of one-loop determinants of chiral fields with $U(1)_R$ charge equal to two and with $U(1)^{\rank \mathfrak{g}}$ charges determined by the non-zero roots $\alpha\neq 0.$

\paragraph{SU(2) gauge theory}
Let us finally then consider the simplest physically relevant example, namely an $SU(2)$ gauge theory with $N_f=N_a=N$ fundamental and antifundamental chiral multiplets. The argument presented above, indicates that factorization can be achieved provided that the R-symmetry is not anomalous, \ie{} if we use the well-known non-anomalous R-charge assignment $r= \frac{N_f-N_c}{N_f}=\frac{N-2}{N}.$ 

The index is computed by
\begin{eqnarray*}
I&=& \frac{1}{2} (p,p)_\infty\ (q,q)_\infty \oint \frac{dz}{2\pi i z} \frac{1}{\Gamma(z^2,p,q)\Gamma(z^{-2}, p,q)} \prod_{\alpha=1}^{N} \Gamma(z^{-1}\zeta_\alpha (pq)^{r/2},p,q)\ \Gamma(z\zeta_\alpha (pq)^{r/2},p,q) \notag\\&& \times \prod_{\beta=1}^{N} \Gamma(z\tilde\zeta_\beta^{-1} (pq)^{r/2}, p,q)\ \Gamma(z^{-1}\tilde\zeta_\beta^{-1} (pq)^{r/2},p,q)\;\notag\\
&=&\frac{1}{2}(p,p)_\infty\ (q,q)_\infty \oint \frac{dz}{2\pi i z} \frac{1}{\Gamma(z^2, p,q)\Gamma(z^{-2}, p,q)}\ \prod_{A=1}^{2N} \Gamma(z^{-1}Y_A, p,q)\ \Gamma(z Y_A,p,q) \;,
\end{eqnarray*}
where we introduced fugacities $\zeta_\alpha, \tilde\zeta_\beta$ for the $SU(N)\times SU(N)$ flavor symmetry. Since the fundamental representation of $SU(2)$ is pseudoreal, we get an enhanced flavor symmetry, with fugacities  $Z_A=(\zeta_\alpha, \tilde\zeta^{-1}_{\beta})\;.$ Finally, we introduced $Y_A=Z_A(pq)^{r/2}.$

The poles from the one factor of the vectormultiplet cancel against the zeros of the other factor and vice versa. 
The integrand further has zeros at $z=p^{-\kappa-1} q^{-\lambda-1} Y_B $ and  $z=p^{\kappa+1} q^{\lambda+1}  Y_C^{-1}$ and poles at $z=p^\kappa q^\lambda Y_B$ and $z= p^{-\kappa} q^{-\lambda} Y_C^{-1}.$ Picking up the poles inside the unit circle, \ie \ the poles at $z=p^\kappa q^\lambda Y_B$, we obtain using the formulas in appendix \ref{section_appendixElliptic}
\begin{eqnarray*}
I&=&\frac{1}{2}\sum_{B=1}^{2N} \frac{ \prod_{A=1}^{2N} \Gamma(Y_B Y_A ; \ p,q) \prod_{\substack{A=1\\A\neq B}}^{2N}\Gamma(Y_B^{-1} Y_A\ ; \ p,q) }{\Gamma(Y_B^2\ ;\ p,q)\Gamma(Y_B^{-2}\ ;\ p,q)}\notag\\
&&\times  \sum_{\kappa,\lambda\geq 0} (pq)^{-2\kappa\lambda(2-N)}\prod_{A=1}^{2N}(Y_A)^{-2\kappa\lambda} \notag\\
&&\times \frac{\prod_{j=1}^{2\lambda}\theta(q^{-j} Y_B^{-2}, p) \prod_{i=1}^{2\kappa}\theta(p^{-i}Y_B^{-2}, q)}{\prod_{j=0}^{2\lambda-1}\theta(q^{j}Y_B^2, p)\prod_{i=0}^{2\kappa-1}\theta(p^{i}Y_B^2, q)}\ \prod_{A=1}^{2N}\frac{\prod_{j=0}^{\lambda-1}\theta(q^{j}Y_AY_B,p) \prod_{i=0}^{\kappa-1}\theta(p^{i}Y_AY_B, q)}{\prod_{j=1}^{\lambda}\theta(q^{-j}Y_B^{-1}Y_A, p)\prod_{i=1}^{\kappa}\theta(p^{-i}Y_B^{-1}Y_A,q)}\;.
\end{eqnarray*}
Note now that the intertwining factor as expected disappears for the correct non-anomalous R-charges: $(pq)^{-2(2-N)}\prod_A(Y_A)^{-2} = (pq)^{-2 ( 2-N + R N )} = 1$ where we used that $\prod_AZ_A=1.$ We thus find complete factorization
\begin{equation}
I = \frac{1}{2}\sum_{B=1}^{2N} Z_{\text{1-loop}}^{\prime (B)}\  Z_{\text{v}}^{(B)} \ Z_{\text{av}}^{(B)},
\end{equation}
where the one-loop contribution is
\begin{equation}
Z_{\text{1-loop}}^{\prime(B)} =  \frac{ \prod_{\substack{A=1\\A\neq B}}^{2N} \Gamma(Y_B Y_A, p,q) \Gamma(Y_B^{-1} Y_A, p,q) }{\Gamma(Y_B^{-2},p,q)}
\end{equation}
and the vortex partition functions are given by
\begin{equation}
Z_{\text{v}}^{(B)} = Z_{\text{vortex}}^{(B)}(p,q,Z_A)\;,\qquad Z_{\text{av}}^{(B)} = Z_{\text{vortex}}^{(B)}(q,p,Z_A)\;.
\end{equation}
Here the vortex membrane partition function is given by
\begin{equation}
Z_{\text{vortex}}^{(B)}(p,q,Z_A) = \sum_{\kappa\geq 0} \frac{\prod_{i=1}^{2\kappa}\theta(p^{-i} Y_B^{-2}, q) }{\prod_{i=\kappa}^{2\kappa-1}\theta(p^{i}Y_B^2, q)} \frac{1}{\prod_{i=1}^{\kappa}\theta(p^{-i}, q)}\prod_{\substack{A=1\\A\neq B}}^{2N}\frac{\prod_{i=0}^{\kappa-1}\theta(p^{i}Y_AY_B,q)}{\prod_{i=1}^{\kappa}\theta(p^{-i}Y_B^{-1}Y_A, q)}\;,
\end{equation}
where $Y_A=Z_A (pq)^{\frac{N-2}{2N}}.$

The generalization of this result to $SU(N)$ gauge group is technically more involved, but is expected to take on a factorized form as well.

\section*{Acknowledgements}
I would like to thank Francesco Benini for collaboration in the initial stages of this project. I am also grateful to Francesco Benini and Leonardo Rastelli for useful discussions and correspondence and to Chris Beem, Jaume Gomis and Shlomo Razamat for comments on the manuscript. I wish to express my gratitude to KITP for its hospitality while this work was being completed. My work is partially supported by NSF Grant PHY-0969919.

\appendix
\section{Spinor conventions}\label{section:appendix_spinorconventions}
We choose to use four-component spinors. Bars on spinors are taken to be the Majorana conjugate, \ie\   $\bar \psi = \psi^t C$ where $C$ is the antisymmetric charge conjugation matrix satisfying $(\gamma_\mu)^{t} C = - C \gamma_\mu$. Since we are in Euclidean signature, it is impossible to impose the Majorana conjugate to be equal to the Dirac conjugate, but rather we work `holomorphically', \ie\ the hermitian conjugate spinor does not make an appearance.

We take the Euclidean gamma matrices to be
\begin{equation}
\gamma^m = \begin{pmatrix}
0 & -i \sigma^m \\
i \bar\sigma^m & 0
\end{pmatrix},
\end{equation}
where $\sigma^m = (\vec\sigma, i\unit_2)$ and $\bar\sigma^m = (\vec\sigma, -i\unit_2),$ where $\vec\sigma$ are the three Pauli matrices. We also introduce $\gamma_5 = \gamma^1\gamma^2\gamma^3\gamma^4=\left(\begin{smallmatrix} \unit_2 & 0 \\ 0 & -\unit_2 \end{smallmatrix}\right)$ which squares to one. The charge conjugation matrix is given explicitly by $C=\gamma^4\gamma^2=\left(\begin{smallmatrix} i\sigma^2 & 0 \\ 0 & -i\sigma^2 \end{smallmatrix}\right).$ 

We also introduce $\sigma^{mn} =  \frac{1}{2}\left( \sigma^m \bar{\sigma}^n - \sigma^n \bar{\sigma}^m \right)$ and $\bar{\sigma}^{mn} =  \frac{1}{2}\left( \bar{\sigma}^m \sigma^n - \bar{\sigma}^n \sigma^m \right),$ in terms of which one can write
\begin{equation}
\gamma_{mn} = \frac{1}{2}\left( \gamma_m \gamma_n - \gamma_n \gamma_m \right) = \begin{pmatrix}
\sigma^{mn} & 0 \\
0& \bar{\sigma}^{mn}
\end{pmatrix}\;.
\end{equation}

Finally, for any four-component spinor $\psi,$ we denote its right and left-handed piece as $\psi_R = \frac{\unit_4 + \gamma_5}{2}\psi$ and $\psi_L = \frac{\unit_4 - \gamma_5}{2}\psi$ respectively.

\section{$\mathcal{N}=1$ supersymmetry algebra on Euclidean four-manifolds}\label{section_appendixalgebra}
In this section we present the $\mathcal{N}=1$ supersymmetry transformation rules on any four-dimensional Euclidean manifold allowing for a solution to the conformal Killing spinor equation $D_\mu \varepsilon = \gamma_\mu \tilde \varepsilon.$  A more general and systematic analysis of supersymmetry on four-dimensional Euclidean backgrounds has been performed in \cite{Festuccia:2011ws,Dumitrescu:2012ha,Dumitrescu:2012at}.

The transformation rules on the vectormultiplet are
\begin{eqnarray}
\delta A_\mu &=& \bar\varepsilon \gamma_\mu \lambda \\
\delta \lambda &=& -\frac{1}{2}\gamma^{\mu\nu} F_{\mu\nu}\ \varepsilon -  \gamma_5 \ D\  \varepsilon \\
\delta D &=& \bar\varepsilon\  \gamma_5 \slashed{D} \lambda\;,
\end{eqnarray}
and those on the chiral multiplet are
\begin{eqnarray}
\delta A &=& \bar\varepsilon \chi \\
\delta B &=& \bar \varepsilon i \gamma_5 \chi \\
\delta \chi &=& \left(\gamma^\mu D_\mu(A+i\gamma_5 B) \right)\varepsilon -i (F+i\gamma_5 G)\varepsilon + \frac{3r}{4}(A-i\gamma_5 B)\slashed{D}\varepsilon \\
\delta F &=& i\bar\varepsilon \slashed{D}\chi + i\left(\frac{3r}{4} - \frac{1}{2} \right)\bar\chi \slashed{D}\varepsilon + \bar\varepsilon (A+i\gamma_5 B) \lambda \\
\delta G &=& -\bar\varepsilon \gamma_5 \slashed{D}\chi +\left(\frac{3r}{4} - \frac{1}{2} \right)\bar\chi \gamma_5\slashed{D}\varepsilon + \bar\varepsilon i\gamma_5(A+i\gamma_5 B) \lambda \; ,
\end{eqnarray}
for commuting $\varepsilon.$ Here $D_\mu$ is the covariant derivative $D_\mu = \partial_\mu - i A_\mu -i V_\mu,$ where $A_\mu$ is the gauge connection and $V_\mu$ is a background field for the R-symmetry. In the chiral multiplet we decomposed $\phi = \frac{A-iB}{2},$ $\bar\phi = \frac{A+iB}{2}$ and $\mathcal{F} = \frac{F+iG}{2},$ $\bar{\mathcal{F}} = \frac{F-iG}{2}.$ The spinor $\varepsilon$ needs to satisfy the Killing spinor equation $D_\mu \varepsilon = \gamma_\mu \tilde \varepsilon.$ One can check that the supersymmetry variations then square to
\begin{equation}\label{algebradeltasq}
\delta^2 = \mathcal{L}^{A+V}_v + \rho \Delta +i \alpha R\;,
\end{equation}
where $\mathcal{L}^{A+V}_v$ is the gauge and background R-symmetry covariant Lie derivative along the vector field $v$, $\Delta$ is the scaling weight\footnote{The scaling weights are 
\begin{eqnarray}
&&\Delta\left(A_\mu, \lambda_R,\lambda_L, D\right)=\left(1,\frac{3}{2},\frac{3}{2},2\right)\;, \qquad \Delta(\varepsilon_R,\varepsilon_L)=\left( \frac{1}{2}, \frac{1}{2}\right)\;, \notag \\
&&\Delta(\phi,\bar\phi,\chi_R,\chi_L, \mathcal{F}, \bar{\mathcal{F}})=\left(\frac{3r}{2},\frac{3r}{2},\frac{3r+1}{2},\frac{3r+1}{2},\frac{3r+2}{2},\frac{3r+2}{2}\right)\;.
\end{eqnarray}
} and $R$ is the $U(1)_R$ generator\footnote{The R-charge assignments are 
\begin{eqnarray}
&&R\left(A_\mu, \lambda_R,\lambda_L, D\right)=\left(0,1,-1,0\right)\;,\qquad R(\varepsilon_R,\varepsilon_L)=\left( 1,-1\right)\;,\notag \\
&& R(\phi,\bar\phi,\chi_R,\chi_L, \mathcal{F}, \bar{\mathcal{F}})=\left(r,-r,r-1,1-r,r-2,2-r\right)\;.
\end{eqnarray}}. The parameters themselves are given by
\begin{equation}\label{algebrabilinears}
v_\mu = \bar\varepsilon \gamma_\mu \varepsilon\,, \qquad \rho = \frac{1}{4}D_\mu v^\mu\,, \qquad \alpha = 3 i \bar{\tilde \varepsilon}\gamma_5 \varepsilon\;.
\end{equation}

\section{Elliptic gamma function}\label{section_appendixElliptic}
The elliptic gamma function is defined as
\begin{equation}
\Gamma(z, p,q) = \prod_{j,k\geq 0}\frac{1-p^{j+1}q^{k+1}/z}{1-p^jq^k z}\;.
\end{equation}
It satisfies the shift formulas
\begin{equation}
\Gamma(pz, p,q) = \theta(z, q) \Gamma(z, p,q)\;, \qquad \Gamma(qz, p,q) = \theta(z,p) \Gamma(z,p,q)\;,
\end{equation}
where $ \theta(z, q) = (z, q)_\infty (q/z, q)_\infty$ in terms of the infinite q-Pochhammer symbol $(z, q)_\infty = \prod_{j\geq 0} (1-zq^j)\;.$ Furthermore, one has
\begin{equation}
\Gamma(z, p,q) \ \Gamma(pq/z,p,q)=1\;.
\end{equation}
The $\theta$-function satisfies
\begin{equation}
\theta(z,q) = \theta(q/z,q) = -z\  \theta(z^{-1}, q)\;,
\end{equation}
which when iterated gives for positive $\kappa$
\begin{equation}
\theta(q^{\kappa} z, q)=\theta\left( z,q\right)\ (-z q^{(\kappa-1)/2})^{-\kappa}, \qquad \theta(q^{-\kappa} z, q)=\theta\left( z,q\right)\ (-z^{-1} q^{(\kappa+1)/2})^{-\kappa}
\end{equation}
Given the above formulae, we can derive for positive $\kappa,\lambda$
\begin{equation}
\Gamma(p^\kappa q^\lambda z, p,q)=\left(-z q^{(\lambda -1)/2} p^{(\kappa -1)/2}\right)^{-\kappa\lambda}\  \Gamma(z, p,q)\ \prod_{j=0}^{\lambda-1} \theta(q^{j} z , p)\ \prod_{i=0}^{\kappa-1}\theta(p^{i} z ,q)\;,
\end{equation}
and
\begin{equation}
\Gamma(p^{-\kappa} q^{-\lambda} z, p,q)= \frac{\Gamma(z, p,q)}{\left(-z^{-1} q^{(\lambda +1)/2} p^{(\kappa +1)/2}\right)^{-\kappa\lambda}\  \prod_{j=1}^{\lambda} \theta(q^{-j} z , p)\ \prod_{i=1}^{\kappa}\theta(p^{-i} z , q)}\;.
\end{equation}

Finally, in order to compute residues, we have the following limit
\begin{equation}
 \lim_{z\rightarrow 1} (1-z)\Gamma(z, p,q) = \frac{1}{(p, p)_\infty\ (q, q)_\infty}\;.
\end{equation} 

%%%%%%%%%%  Bibliography  %%%%%%%%%%%%
{%\small
\bibliographystyle{utphys}
\bibliography{4dHiggsLocalization}
}

\end{document}